\documentclass[11pt]{article}
\title
{Disproving the Neighbourhood Conjecture}
\author{Heidi Gebauer
\thanks{Institute of
Theoretical Computer Science, ETH Zurich, CH-8092 Switzerland. Email:
gebauerh@inf.ethz.ch. } 
}

\usepackage{amsmath, amsfonts}
\usepackage{subfigure}
\usepackage[dvips]{graphicx} 

\newcommand{\fref}[1]{Figure~\ref{#1}}

\begin{document}
\bibliographystyle{plain}
\maketitle
\newtheorem{theo}{Theorem} [section]
\newtheorem{defi}[theo]{Definition}
\newtheorem{lemm}[theo]{Lemma}                                                                                                                                                                                                                                                                                                                                                                                                                                                                                                                   
\newtheorem{obse}[theo]{Observation}
\newtheorem{prop}[theo]{Proposition}
\newtheorem{coro}[theo]{Corollary}
\newtheorem{rem}[theo]{Remark}

\newcommand{\whp}{{\bf whp}}
\newcommand{\prob}{probability}
\newcommand{\rn}{random}
\newcommand{\rv}{random variable}
\newcommand{\hpg}{hypergraph}
\newcommand{\hpgs}{hypergraphs}
\newcommand{\subhpg}{subhypergraph}
\newcommand{\subhpgs}{subhypergraphs}
\newcommand{\bH}{{\bf H}}
\newcommand{\cH}{{\cal H}}
\newcommand{\cT}{{\cal T}}
\newcommand{\cF}{{\cal F}}
\newcommand{\cG}{{\cal G}}
\newcommand{\cD}{{\cal D}}
\newcommand{\cC}{{\cal C}}

\newcommand{\ideg}{\mathsf {ideg}}
\newcommand{\lv}{\mathsf {lv}}
\newcommand{\nga}{n_{\text{game}}}
\newcommand{\avdaneg}{\overline{\deg}}
\newcommand{\ed}{e_{\text{double}}}

\newcommand{\danger}{\mathsf {dang}}
\newcommand{\avdanan}{\overline{\danger}}

\newcommand{\degb}{\deg_{B}}
\newcommand{\degm}{\deg_{M}}

\newcommand{\avd}{\overline{D}}

\begin{abstract}
We study the following Maker/Breaker game.
Maker and Breaker take turns in choosing vertices from a given $n$-uniform hypergraph $\cal{F}$, with Maker going first.
Maker's goal is to completely occupy a hyperedge and Breaker tries to avoid this. 
Beck conjectures that if the maximum neighborhood size of $\cal{F}$ is at most $2^{n - 1}$ then Breaker has a winning strategy. We disprove this conjecture by establishing an $n$-uniform hypergraph with maximum neighborhood size $3 \cdot 2 ^{n - 3}$ where Maker has a winning strategy. Moreover, we show how to construct an $n$-uniform hypergraph  with maximum degree $\frac{2^{n - 1}}{n}$ where Maker has a winning strategy.

Finally we show that each $n$-uniform hypergraph with maximum degree at most $\frac{2^{n - 2}}{en}$ has a proper halving 2-coloring, which solves 
another open problem posed by Beck related to the Neighbourhood Conjecture.

\end{abstract}

\section{Introduction}

A \emph{hypergraph} is a pair $(V,E)$, where $V$ is a finite set whose elements are called \emph{vertices} and $E$ is a family of subsets of $V$, called \emph{hyperedges}. 
We study the following Maker/Breaker game.
Maker and Breaker take turns in claiming 
one previously unclaimed vertex of a given $n$-uniform hypergraph, with Maker going first.
Maker wins if he claims all vertices of some hyperedge of $\cal{F}$, otherwise Breaker wins.

Let $\cal{F}$ be a $n$-uniform hypergraph. The \emph{degree} $d(v)$ of a vertex $v$ is the number of hyperedges containing $v$ and the \emph{maximum degree} of $\cal{F}$ is the maximum degree of its vertices. The \emph{neighborhood} $N(e)$ of a hyperedge $e$ is the set of hyperedges of $\cal{F}$ which intersect $e$ and the \emph{maximum neighborhood size} of $\cal{F}$ is the maximum of $|N(e)|$ where $e$ runs over all hyperedges of $\cal{F}$.

The famous Erd\H{o}s-Selfridge Theorem \cite{ES} states that for each $n$-uniform hypergraph $\cal{F}$ with less than $2 ^{n - 1}$ hyperedges Breaker has a winning strategy. This upper bound on the number of hyperedges is best possible as the following example shows. 
Let $T$ be a rooted binary tree with $n$ levels and let $\cal{G}$ be the hypergraph whose hyperedges are exactly the sets $\{v_{0}, \ldots v_{n - 1}\}$ such that $v_{0}, v_{1}, \ldots, v_{n - 1}$ is a path from the root to a leaf.
Note that the number of hyperedges of $\cal{G}$ is $2^{n - 1}$. 
To win the game on $\cal{G}$ Maker can use the following strategy. In his first move he claims the root $m_{1}$ of $T$. Let $b_{1}$ denote the vertex occupied by Breaker in his subsequent move. 
In his second move Maker claims the child $m_{2}$ of $m_{1}$ such that $m_{2}$ lies in the subtree of $m_{1}$ not containing $b_{1}$. 
More generally, in his $i$th move Maker selects the child $m_{i}$ of his previously occupied node $m_{i - 1}$ such that the subtree rooted at $m_{i}$ contains no Breaker's node.
Note that such a child $m_{i}$ always exists since the vertex previously claimed by Breaker is either in the left or in the right subtree of $m_{i-1}$ (but not in both!). Using this strategy Maker can achieve to own some set $\{v_{0}, \ldots, v_{n - 1}\}$ of vertices such that $v_{0}, v_{1}, \ldots, v_{n - 1}$ is a path from the root to a leaf, which corresponds to some hyperedge of $\cal{G}$. Hence Maker has a winning strategy on $\cal{G}$.

Note that both the maximum neighborhood size and the maximum degree of $\cal{G}$ are $2^{n - 1}$, thus equally large as the number of hyperedges of $\cal{G}$. 
This provides some evidence that in order to be a Maker's win a hypergraph must have largely overlapping hyperedges. Moreover, Beck \cite{B} conjectured that the main criterion for whether a hypergraph is a Breaker's win is not the cardinality of the hyperedge set but rather the maximum neighborhood size, i.e. the actual reason why each hypergraph $\cal{H}$ with less than $2^{n-1}$ edges is a Breaker's win is that the maximum neighborhood of $\cal{H}$ is smaller than $2^{n-1}$.
%
\newline
\newline
\textbf{Neighborhood Conjecture} (Open Problem 9.1(a), \cite{B}) Assume that $\cal{F}$ is an $n$-uniform \hpg, and its maximum neighborhood size is smaller than $2^{n - 1}$. Is it true that by playing on $\cal{F}$ Breaker has a winning strategy?
\newline
\newline
Further motivation for the Neighborhood Conjecture is the well-known Erd\H{o}s-Lov\'{a}sz 2-coloring Theorem -- a direct consequence of the famous Lov\'{a}sz Local Lemma -- which states that every $n$-uniform hypergraph with maximum neighborhood size at most $2^{n-3}$ has a proper 2-coloring. 
An interesting feature of this theorem is that the board size does not matter.
In this paper we prove by applying again the Lov\'{a}sz Local Lemma that in addition every  $n$-uniform hypergraph with maximum neighborhood size at most $\frac{2^{n-3}}{n}$ has a so called \emph{proper halving} 2-coloring, i.e., a proper 2-coloring in which the number of red vertices and the number of blue vertices differ by at most 1 (see Theorem \ref{theo:NeighbourhoodPartProperHalvingIsOk} for details). This guarantees the existence of a course of the game at whose end Breaker owns at least one vertex of each hyperedge and thus is the winner. This suggests that the game we study is a priori not completely hopeless for Breaker.


In our first theorem we prove that the Neighborhood Conjecture, in this strongest of its forms, is not true.

\begin{theo} \label{theo:NeighbourhoodPartanotwork}
There is an $n$-uniform hypergraph $\cal{H}$ with maximum neighborhood size $2^{n-2} + 2^{n - 3}$ where Maker has a winning strategy
\end{theo}
In the hypergraph $\cal{H}$ we will construct to prove Theorem \ref{theo:NeighbourhoodPartanotwork}
one vertex has degree $2^{n-2}$. However, the existence of vertices with high degree is not crucial. 
We can also establish a hypergraph with maximum degree $\frac{2^{n-1}}{n}$ on which Maker has a winning strategy. In this case the maximum neighborhood size is at most $2^{n - 1} - n$, which is weaker than Theorem \ref{theo:NeighbourhoodPartanotwork} but also disproving the Neighborhood Conjecture.

\begin{theo} \label{theo:NeighbourhoodPartbnotworkc1}
There is an $n$-uniform hypergraph $\cal{H}$ with maximum degree $\frac{2^{n-1}}{n}$ where Maker has a winning strategy.
\end{theo}

In his book \cite{B} Beck also poses several weakenings of the Neighborhood Conjecture, i.e. 

\begin{itemize}
\item[(i)] (Open Problem 9.1(b), \cite{B}) If the Neighborhood Conjecture is too difficult (or false) then how about if the upper bound on the maximum neighborhood size is replaced by an upper bound $\frac{2^{n - c}}{n}$ on the maximum degree where $c$ is a sufficiently large constant?

\item[(ii)] (Open Problem 9.1(c), \cite{B}) If (i) is still too difficult, then how about a polynomially weaker version where the upper bound on the maximum degree is replaced by $n^{-c} \cdot 2^{n}$, where $c > 1$ is a positive absolute constant?

\item[(iii)] (Open Problem 9.1(d), \cite{B}) If (ii) is still too difficult, then how about an exponentially weaker version where the upper bound on the maximum degree is replaced by $c^{n}$, where $2 > c > 1$ is an absolute constant?

\item[(iv)] (Open Problem 9.1(e), \cite{B}) How about if we make the assumption that the hypergraph is almost disjoint?

\item[(v)] (Open Problem 9.1(f), \cite{B}) How about if we just want a proper halving 2-coloring?

\end{itemize}
Note that Theorem \ref{theo:NeighbourhoodPartbnotworkc1} disproves (i) for $c = 1$. 
%
%

Finally we deal with (v). It is already known that the answer is positive if the maximum degree is at most $\left(\frac{3}{2} - o(1)\right)^{n}$. According to Beck \cite{B} the real question in (v) is whether or not $\frac{3}{2}$ can be replaced by 2. We prove that the answer is yes.

\begin{theo} \label{theo:NeighbourhoodPartProperHalvingIsOk}
For every $n$-uniform hypergraph $\cal{F}$ with maximum degree at most $\frac{2^{n - 2}}{en}$ there is a proper halving 2-coloring.
\end{theo}
Before starting with the actual proofs we fix some notation. 
Let $T$ be a rooted binary tree of height $h$. With a \emph{path} of $T$ we denote an ordinary path $v_{i}, v_{i + 1}, \ldots, v_{j}$ of $T$ where $v_{k}$ is on level $k$ for every $k = i, \ldots, j$. A \emph{branch} of $T$ is a path starting at the root of $T$. Finally, a \emph{full branch} of $T$ is a branch of length $h + 1$. 
The hypergraphs we will construct to prove Theorem \ref{theo:NeighbourhoodPartanotwork} and Theorem \ref{theo:NeighbourhoodPartbnotworkc1} both belong to the class $\cal{C}$ of hypergraphs $\cal{H}$ whose vertices can be arranged in a binary tree $T_{\cal{H}}$ such that each hyperedge of $\cal{H}$ is a path of $T_{\cal{H}}$.
Depending on the context we consider a hyperedge $e$ of a hypergraph $\cal{H}$ either as a set or as a path in $T_{\cal{H}}$. So we will sometimes speak of the start or end node of a hyperedge. 

\section{Counterexample to the Neighborhood Conjecture}

\begin{figure} [!htb]
\centering
\includegraphics[width=0.7\textwidth]{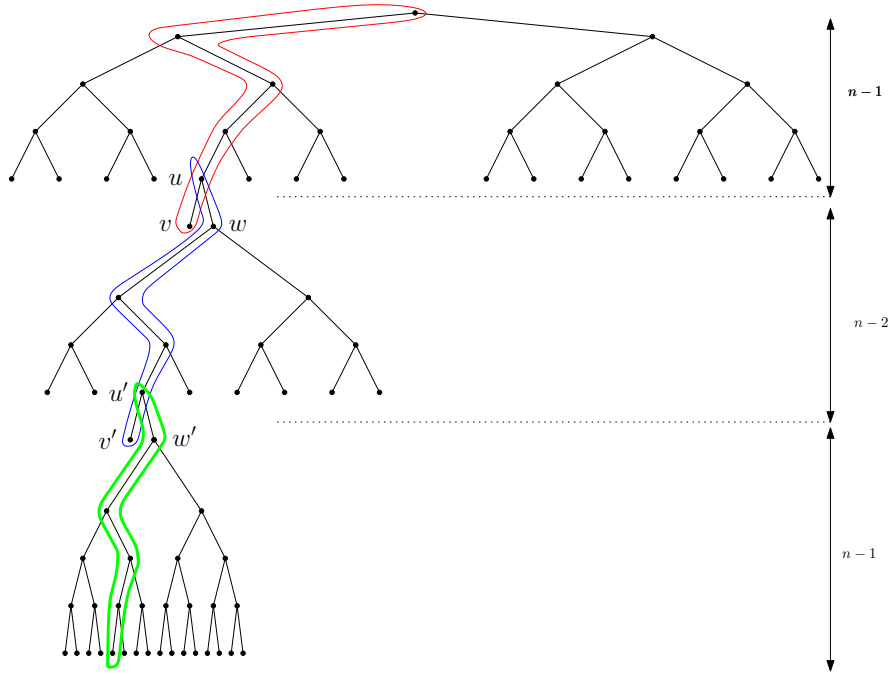}
\caption{An illustration of $\cal{H}$. The marked paths represent exemplary hyperedges} \label{fig:DisprovingNC}.
\end{figure}

\emph{Proof of Theorem \ref{theo:NeighbourhoodPartanotwork}:} Our goal is to construct an element $\cal{H} \in \cal{C}$ with the required maximum neighborhood size where Maker has a winning strategy. Before specifying $\cal{H}$ we fix Maker's strategy.
In his first move he claims the root $m_{1}$ of $T_{\cal{H}}$.
In his $i$th move he then selects the child $m_{i}$ of his previously occupied node $m_{i - 1}$ such that the subtree rooted at $m_{i}$ contains no Breaker's vertex.
Note that such a child $m_{i}$ always exists since the vertex previously claimed by Breaker is either in the left or in the right subtree of $m_{i-1}$ (but not in both!). 
This way Maker can achieve some full branch of $T_{\cal{H}}$ by the end of the game. This directly implies the following. 
\begin{obse} \label{obse:NeighborhoodGeneralObservationBranch}
Let $\cal{G} \in \cC$ be an $n$-uniform hypergraph such that every full branch of $T_{\cal{G}}$ contains a hyperedge. Then Maker has a winning strategy on $\cal{G}$.
\end{obse}
So in order to prove Theorem \ref{theo:NeighbourhoodPartanotwork} it suffices to show the following claim.
\begin{lemm} \label{lemm:NeighbourhoodPartaAuxilLemm}
There is an $n$-uniform hypergraph $\cal{H} \in \cal{C}$ with maximum neighborhood $2^{n - 2} + 2^{n - 3}$ such that each full branch of $T_{\cH}$ contains a hyperedge of $\cH$.
 \end{lemm} $\Box$
 \newline
\emph{Proof of Lemma \ref{lemm:NeighbourhoodPartaAuxilLemm}:}
We construct 
$\cal{H}$ as follows. Let $T'$ be a binary tree with $n - 1$ levels. For each leaf $u$ of $T'$ we proceed as follows. Then we add two children $v$, $w$ to $u$ and let the full branch ending at $v$ be a hyperedge.
Then we attach a subtree $S$ with $n-2$ levels to $w$ (such that $w$ is the root of $S$). 
We need to achieve that each full branch containing $w$ contains a hyperedge.
For each leaf $u'$ of $S$ we therefore do the following. 
We add two children $v'$, $w'$ to $u'$ and let the path from $u$ to $v'$ be a hyperedge. Moreover, we attach a subtree $S'$ with $n - 1$ levels to $w'$ (such that $w'$ is the root of $S'$).
We have to complete our tree in such a way that each full branch containing $w'$ contains a hyperedge. To this end we let each path from $u'$ to a leaf of $S'$ be a hyperedge. \fref{fig:DisprovingNC} shows an illustration. 
It remains to show that the maximum neighborhood of the resulting 
hypergraph $\cal{H}$ is at most $2^{n - 2} + 2^{n - 3}$.
\begin{prop} \label{prop:Neighborhoodnumberintersections}
 Every hyperdge $e$ of $\cal{H}$ intersects at most $2^{n - 2} + 2^{n - 3}$ other hyperdges.
\end{prop}
$\Box$
\newline
\emph{Proof of Proposition \ref{prop:Neighborhoodnumberintersections}:} 
We fix six vertices $u, u', v, v', w, w'$ according to the above description, i.e., $u$ is a node on level $n -2$ whose children are $v$ and $w$, $u'$ is a descendant of $w$ on level $2n - 4$ whose children are $v'$ and $w'$.
Let $e$ be a hyperedge of $\cal{H}$. Note that the start node of $e$ is either the root $r$ of $T_{\cal{H}}$, a node on the same level as $u$ or a node on the same level as $u'$. We now distinguish these cases.
\begin{itemize}
\item[\textbf{Case (a)}:] The start node of $e$ is $r$.
\newline
By symmetry we assume that $e$ ends at $v$. According to the construction of $T_{\cal{H}}$ the hyperedge $e$ intersects the $2^{n - 2} - 1$ other hyperedges starting at $r$ and the $2^{n - 3}$ hyperedges starting at $u$. So altogether $e$ intersects  $2^{n - 2} + 2^{n - 3} - 1$ hyperedges, as claimed.
\item[\textbf{Case (b)}:] The start node of $e$ is on the same level as $u$.
\newline
By symmetry we suppose that $e$ starts at $u$ and ends at $v'$. 
The hyperdges intersecting $e$ can be divided into the following three categories.
\begin{itemize}
\item The hyperedge starting at $r$ and ending at $v$,
\item the $2^{n - 3} - 1$  hyperedges different from $e$ starting at $u$, \enspace and
\item the $2^{n - 2}$ hyperedges starting at $u'$,
\end{itemize}
implying that $e$ intersects at most $2^{n - 2} + 2^{n - 3}$ hyperedges in total.
\item[\textbf{Case (c)}:] The start node of $e$ is on the same level as $u'$
\newline
By symmetry we assume that $e$ starts at $u'$. Then $e$ intersects the $2^{n - 2}$ other hyperedges starting at $u'$ and the hyperedge starting at $u$ and ending at $v'$, thus $2^{n - 2} + 1$ hyperedges altogether.
\end{itemize}
$\Box$

\section{A Degree-Regular hypergraph with small maximum degree which is a Maker's win.}  

We need some notation first.
Throughout this paper $\log$ will denote logarithm to the base 2.
The vertex set and the hyperedge set of a {\hpg} $\cal{G}$ are denoted by $V(\cal{G})$ and $E(\cal{G})$, respectively. By a slight abuse of notation we consider $E(\cal{G})$ as a multiset, i.e. each hyperedge $e$ can have a multiplicity greater than 1.
By a \emph{bottom hyperedge} of a tree $T_{\cG}$ we denote a hyperedge covering a leaf of $T_{\cal{G}}$. 
As in the previous section we only deal with hypergraphs of the class $\cal{C}$.
 
Before tackling the rather technical proof of Theorem \ref{theo:NeighbourhoodPartbnotworkc1} we show the following weaker claim.
 
\subsection{A weaker statement}

\begin{figure} [!htb]
\centering
\includegraphics[width=0.3\textwidth]{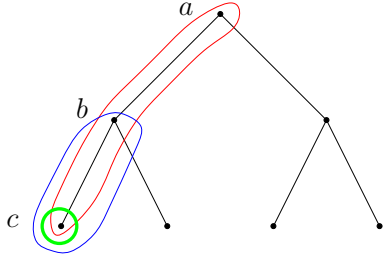}
\caption{An illustration of $\cal{H}$ for $d = 4$. The hyperedge $\{a,b,c\}$ has multiplicity 1,  $\{b,c\}$ has multiplicity 2 and $\{c\}$ has multiplicity 4.} \label{fig:WeakerClaim}
\end{figure}

\begin{theo} \label{theo:NeighbourhoodPartbweakerclaim}
There is a $n$-uniform hypergraph $\cal{H}$ with maximum degree $\frac{2^{n+1}}{n}$ where Maker has a winning strategy.
\end{theo}
 Let $d = \frac{2^{n}}{n}$. For simplicity we assume that $n$ is a power of 2, implying that $d$ is power of 2 as well.
Due to Observation  \ref{obse:NeighborhoodGeneralObservationBranch} it suffices to show the following.
\begin{lemm} \label{lemm:NeighbourhoodPartbweakerH}
 There is an $n$-uniform hypergraph $\cal{G} \in \cal{C}$ with maximum degree $2d$ such that every full branch of $T_{\cG}$ contains a hyperedge of $\cG$.
 \end{lemm} $\Box$
\newline
\emph{Proof of Lemma \ref{lemm:NeighbourhoodPartbweakerH}:}
To construct the required hypergraph $\cal{G}$ 
we establish first a (not necessarily $n$-uniform) hypergraph $\cal{H}$ and then successively modify its hyperedges and $T_{\cal{H}}$.
The following lemma is about the first step.
\begin{lemm} \label{lemm:First}
There is a hypergraph $\cal{H} \in \cal{C}$ with maximum degree $2d$ such that every full branch of $T_{\cal{H}}$ has $2^{i}$ bottom hyperedges of size $\log d + 1 - i$ for every $i$ with $0 \leq i \leq \log d$.
 \end{lemm} 
\emph{Proof of Lemma \ref{lemm:First}:} 
Let $T$ be a binary tree with $\log d + 1$ levels. In order to construct the desired hypergraph $\cal{H}$ we proceed for each vertex $v$ of $T$ as follows. 
For each leaf descendant $w$ of $v$ we let the path from $v$ to $w$ be a hyperedge of multiplicity $2^{l(v)}$ where $l(v)$ denotes the level of $v$.
Figure \ref{fig:WeakerClaim} shows an illustration.
The construction yields that each full branch of $T_{\cal{H}}$ has $2^{i}$ bottom hyperedges of size $\log d + 1 - i$ for every $i$ with $0 \leq i \leq \log d$. So it remains to show that $d(v) \leq 2d$ for every vertex of $v \in V(T)$. 
Note that every vertex $v$ has $2^{\log d - l(v)}$ leaf descendants in $T_{\cal{H}}$, implying that $v$ is the start node of  $2^{\log d - l(v)} \cdot 2^{l(v)} \leq d$ hyperedges. So the degree of the root is at most $d \leq 2d$. We then apply induction. Suppose that $d(u) \leq 2d$ for all nodes $u$ with $l(u) \leq i - 1$ for some $i$ with $1 \leq i \leq \log d$ and let $v$ be a vertex on level $i$. By construction exactly half of the hyperedges containing the ancestor of $v$ also contain $v$ itself. Hence $v$ occurs in at most $\frac{1}{2} \cdot 2d = d$ hyperedges as non-start node. Together with the fact that $v$ is the start node of at most $d$ hyperedges this implies that $d(v) \leq d + d \leq 2d$. $\Box$

The next lemma deals with the second step of the construction of the required hypergraph  $\cal{G}$.
\begin{lemm} \label{lemm:Second}
 There is a hypergraph $\cal{H'} \in \cal{C}$ with maximum degree $2d$ such that each full branch of $T_{\cal{H'}}$ has $2^{i}$ bottom hyperedges of size $\log d + 1 - i + \lfloor \log \log d \rfloor$ for some $i$ with $0 \leq i \leq \log d$.
 \end{lemm} 
\emph{Proof:} 
Let $\cal{H} \in \cal{C}$ be a hypergraph with maximum degree $2d$ such that every leaf $u$ of $T_{\cal{H}}$ is the end node of a set $S_{i}(u)$ of $2^{i}$ hyperedges of size $\log d + 1 - i$ for every $i$ with $0 \leq i \leq \log d$. (Lemma \ref{lemm:First} guarantees the existence of $\cal{H}$.) To each leaf $u$ of $T_{\cal{H}}$ we then attach a binary tree $T'_{u}$ of height $\lfloor \log \log d \rfloor$ in such a way that $u$ is the root of $T'_{u}$. 
Let $v_{0}, \ldots, v_{2^{\lfloor \log \log d \rfloor} - 1}$ denote the leaves of $T'_{u}$. For every $i$ with $0 \leq i \leq 2^{\lfloor \log \log d \rfloor} - 1$ we then augment every hyperedge of $S_{i}(u)$ with the set of vertices different from $u$ along the full branch of $T'_{u}$ ending at $v_{i}$.
 
After repeating this procedure for every leaf $u$ of $T_{\cal{H}}$ we get the desired hypergraph $\cal{H'}$. It remains to show that every vertex in $\cal{H'}$ has degree at most $2d$. To this end note first that during our construction the vertices of $\cal{H}$ did not change their degree. Secondly, let $u$ be a leaf of $T_{\cal{H}}$. By assumption $u$ has degree at most $2d$ and by construction $d(v) \leq d(u)$ for all vertices $v \in V(\cal{H'}) \backslash V(\cal{H})$, which completes our proof. $\Box$

\begin{lemm} \label{lemm:NThird}
There is a hypergraph $\cal{H''} \in \cal{C}$ with maximum degree $2d$ such that every full branch of $T_{\cal{H''}}$ has one bottom hyperedge of size $\log d + 1  + \lfloor \log \log d \rfloor$.
 \end{lemm} 
Note that due to our choice of $d$, Lemma \ref{lemm:NThird} directly implies Lemma \ref{lemm:NeighbourhoodPartbweakerH}. $\Box$
 \newline
 \emph{Proof of Lemma \ref{lemm:NThird}:} By Lemma \ref{lemm:Second} there is a hypergraph $\cal{H'} \in \cal{C}$ with maximum degree $2d$ such that each full branch of $T_{\cal{H'}}$ has $2^{i}$ bottom hyperedges of size $\log d + 1 - i + \lfloor \log \log d \rfloor$ for some $i$ with $0 \leq i \leq \log d$. For every leaf $u$ of $T_{\cal{H'}}$ we proceed as follows. Let $e_{1}, \ldots, e_{2^{i}}$ denote the bottom hyperedges of $\cal{H'}$ ending at $u$. We then attach a binary tree $T''$ of height $i$ to $u$ in such a way that $u$ is the root of $T''$. 
Let $p_{1}, \ldots, p_{2^{i}}$ denote the full branches of $T''$. We finally augment $e_{j}$ with the vertices along $p_{j}$, for $j = 1 \ldots 2 ^{i}$.

After repeating this procedure for every leaf $u$ of $T_{\cal{H'}}$ we get the resulting graph $\cal{H''}$.
By construction every full path of $T_{\cal{H''}}$ has one bottom hyperedge of size $\log d + 1  + \lfloor \log \log d \rfloor$.
A similar argument as in the proof of Lemma \ref{lemm:Second} shows that the maximum degree of $\cal{H''}$ is at most $2d$. $\Box$
 \newline
To prove Theorem \ref{theo:NeighbourhoodPartbnotworkc1} we then use the same basic ideas, augmented with some refined analysis. To achieve the additional factor of $\frac{1}{4}$ in the bound on the maximum degree we however have to deal with many technical issues. 

\subsection{The actual Theorem}


We fix some notation first.
A \emph{unit} is a set of $2^{i}$ hyperedges of size $\log d + 1 - i$ for some $i \leq \log(d) + 1$. Similarly, a \emph{unit} of \emph{power} $k$ denotes a set of $2^{i}$ hyperedges of size 
$\log d + 1 - i + k$ for some $i \leq \log(d) + 1$. Let $U$ be a unit. By a slight abuse of notation we let the \emph{length} $l(U)$ of a unit $U$ denote the size of the hyperedges of $U$. 
Accordingly, a unit is called a \emph{bottom unit} if all of its hyperedges are bottom hyperedges.

Note that we have already used the term of a unit  implicitly in the proof of Theorem \ref{theo:NeighbourhoodPartbweakerclaim}, e.g. the hypergraph $\cal{H}$ mentioned in Lemma \ref{lemm:First} has the property that each full branch of $T_{\cal{H}}$ has $\log d + 1$ bottom units of length at most $\log d + 1$ each, the hypergraph $\cal{H'}$ 
of Lemma \ref{lemm:Second} corresponds to a tree $T_{\cal{H'}}$ where each full branch contains one bottom unit of power $\lfloor \log \log d \rfloor$ and, finally, in the tree $T_{H''}$ of Lemma \ref{lemm:NThird} every full branch contains a bottom unit of length $n$, which represents an ordinary hyperedge of size $n$. 

\emph{Proof of Theorem \ref{theo:NeighbourhoodPartbnotworkc1}:}  
Due to Observation \ref{obse:NeighborhoodGeneralObservationBranch} it suffices to show the following.
\begin{lemm} \label{lemm:NeighbourhoodPartbconstructH}
 There is an $n$-uniform hypergraph $\cal{H} \in \cal{C}$ with maximum degree $\frac{2^{n - 1}}{n}$ such that every full branch of $T_{\cH}$ contains a hyperedge of $\cH$.
 \end{lemm} $\Box$
\newline
\emph{Proof of Lemma \ref{lemm:NeighbourhoodPartbconstructH}}

Let $d = \frac{2^{n - 2}}{n}$. For simplicity we assume that $n$ is a power of 2, implying that $d$ is a power of 2. From now on by a \emph{hypergraph} we mean an ordinary hypergraph of $\cal{C}$ with maximum degree $2d$.

We now state some technical lemmas.

\subsubsection{General Facts}

\begin{figure} [!htb]
\centering
\includegraphics[width=0.35\textwidth]{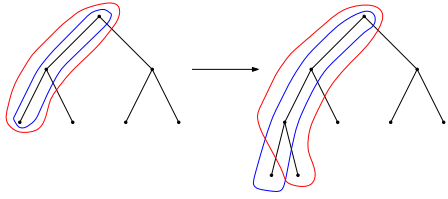}
\caption{Splitting a node.} \label{fig:RegularConstruction}
\end{figure}

The basic operation we use in our construction will be denoted by \emph{node splitting}. Let $\cal{G}$ be a hypergraph and let $u$ be a leaf of $T_{\cal{G}}$ such that there is a set $S$ of bottom hyperedges ending at $u$. Then \emph{splitting} $u$ means that we add two children $v_{1}, v_{2}$ to $u$, partition $S$ into two subsets $S_{1}, S_{2}$ and augment every hyperedge of $S_{i}$ with $v_{i}$ for $i = 1,2$. Possibly we also add new hyperedges of size 1 containing either $v_{1}$ or $v_{2}$. \fref{fig:RegularConstruction} shows an illustration for $|S| = 2$. 
We will often apply a series of hyperedge splittings.
By \emph{extending} a hypergraph $\cG$ \emph{at} a leaf $u$ of $T_{\cG}$ we denote the process of successively splitting one of the current leaves in the subtree of $u$;
i.e., the resulting hypergraph can be obtained by adding to $u$ a left and a right subtree, modifying the hyperedges of $\cal{G}$ containing $u$ and possibly adding some new hyperedges starting at a descendant of $u$ (the other hyperedges remain as they are).

The next lemma is about another basic modification.
\begin{lemm} \label{lemm:NeighbourhoodbMoreGeneralizedUnitsIncrease}
Let $\cG$ be a hypergraph and let $u$ be a leaf of $T_{\cG}$ such that the full branch of $T_{\cG}$ ending at $u$ contains $i$ bottom units $U_{1}, \ldots, U_{i}$ with $l(U_{j}) \leq \log d$. Then $u$ can be  split in such a way that each full branch containing $u$ has $i + 1$ bottom units $U'_{1}, \ldots, U'_{i + 1}$ with $l(U'_{1}) = 1$ and $l(U'_{j + 1}) = l(U_{j}) + 1$ for $j = 1 \ldots i$.
\end{lemm}
\emph{Proof:} Let $v_{1}$, $v_{2}$ be the children of $u$. For each $U_{i}$ we proceed as follows. To half of the hyperedges of $U_{i}$ we add $v_{1}$ and to the other half we add $v_{2}$. Finally, we let $\{v_{1}\}$, $\{v_{2}\}$ be hyperedges occurring with multiplicity $d$ each. Let $\cal{G'}$ denote the resulting hypergraph. By construction $\cal{G'}$ fulfills the requirements of Lemma \ref{lemm:NeighbourhoodbMoreGeneralizedUnitsIncrease} as far as the bottom units $U'_{1}, \ldots, U'_{i + 1}$ are concerned. It remains to show that $\cal{G}$ has maximum degree $2d$.
To this end note that apart from $v_{1}$ and $v_{2}$ all vertices of $\cal{G'}$ have the same degree as in $\cal{G}$. The construction yields that $d_{\cal{G'}}(v_{1}), d_{\cal{G'}}(v_{2}) \leq d + \frac{d_{\cG}(u)}{2}$. Since by assumption $d_{\cG}(u) \leq 2d$ we are done. $\Box$

Note that Lemma \ref{lemm:First} states that there is a hypergraph $\cal{H} \in \cal{C}$ such that each full branch of $T_{\cal{H}}$ has $\log d + 1$ bottom units of length at most $\log d + 1$. We generalize this fact in the following two statements, which are both direct Corollaries of Lemma \ref{lemm:NeighbourhoodbMoreGeneralizedUnitsIncrease}.

\begin{coro} \label{coro:NeighbourhoodbAugmentUnits}
Let $i \leq \log d + 1$. 
Then there is a hypergraph $\cG$ such that each full branch of $T_{\cG}$ contains  $i$ bottom units $U_{1}, \ldots, U_{i}$ with $l(U_{j}) = j$ for $j = 1 \ldots i$.
\end{coro}
\begin{coro} \label{coro:NeighbourhoodbMoreGeneralizedUnitsIncrease}
Let $r \leq s$ be integers with $s \leq \log d + 1$.
Let $\cG$ be a hypergraph and let $u$ be a leaf of $T_{\cG}$ such that the full branch ending at $u$ contains $i$ bottom units $U_{1}, \ldots, U_{i}$ with $l(U_{j}) \leq r$ for every $j = 1, \ldots, i$. Then $\cal{G}$ can be extended at $u$ in such a way that in the tree $T_{\cal{G'}}$ corresponding to the resulting hypergraph $\cal{G'}$ each full branch containing $u$ has $i + s - r$ bottom units $V_{1}, \ldots, V_{s - r}$, $V'_{1}, \ldots, V'_{i}$ with $l(V_{j}) = j$ for $j = 1 \ldots s - r$  and $l(V'_{j}) = l(U_{j}) + s - r$ for $j = 1 \ldots i$.
\end{coro}
Next we describe how one can develop some units by giving up others. 
Let $k \geq 0$ and let $i$ be an even number.
Suppose there is a hypergraph $\cG$ and a vertex $u \in V(\cal{G})$ such that $u$ is a leaf 
of $T_{\cal{G}}$ and the full branch ending at $u$ contains $i$ bottom units $U_{1}, \ldots, U_{i}$ of power $k$ each.  
Then $u$ can be split in such a way that each full branch of containing $u$ has $\frac{i}{2}$ bottom units of power $k + 1$. Indeed, we just have to split $u$ in such a way that one child $v$ of $u$ is added to all hyperedges of $U_{j}$ for every $j \leq \frac{i}{2}$ whereas the other child $w$ of $u$ is added to all hyperedges of $U_{j}$ for every $j \geq \frac{i}{2} + 1$. This directly implies the following.

\begin{prop} \label{prop:NeighbourhoodbMoreGeneralizedUnitsShrink}
Let $k \geq 0$ and let $i$ be a power of 2.
Suppose that there is a hypergraph $\cG$ and a leaf $u$ of $T_{\cal{G}}$ such that the full branch ending at $u$ contains $i$ bottom units $U_{1}, \ldots, U_{i}$ of power $k$ each. Then $\cal{G'}$ can be extended at $u$ in such a way that in the tree $T_{\cal{G'}}$ of the resulting hypergraph $\cal{G'}$ each full branch containing $u$ has a bottom unit of power $k + \log i$.
\end{prop}
We describe some other frequently applied modifications of hypergraphs. 
Let $k \geq 0$, let $\cG$ be a hypergraph and let $u$ be a leaf of $T_{\cal{G}}$ such that the full branch ending at $u$ contains a bottom unit $U$ of power $k$ with $|U| \geq 2$. Similarly as above we can split $u$ in such a way that each full branch containing $u$ has a bottom unit $U'$ of power $k$ with $|U'| = \frac{|U|}{2}$. 
By successively splitting the descendants of $u$ in this way we obtain that finally (in the resulting tree) each full branch containing $u$ has a bottom unit of power $k$ with $|U| = 1$. Together with the fact that a unit $U$ of power $k$ with $|U| = 1$ must have length $\log d + k + 1$ this implies that to show Lemma \ref{lemm:NeighbourhoodPartbconstructH} it is sufficient to establish a hypergraph $\cG$ where each full branch of $T_{\cG}$ contains one bottom unit of power $n - \log d - 1$. Together with Proposition \ref{prop:NeighbourhoodbMoreGeneralizedUnitsShrink} this implies the following.
\begin{obse} \label{obse:NeighbourhoodbMoreGeneralizedsufficientconditiontowin}
Suppose that there is a hypergraph $\cG$ where each full branch $P$ of $T_{\cG}$ contains $l_{P}$ bottom units of power $k_{P}$ such that $k_{P} + \lfloor \log l_{P} \rfloor \geq n - \log d - 1$. Then Lemma \ref{lemm:NeighbourhoodPartbconstructH} holds.
\end{obse}
We are now able to roughly describe the actual construction of $\cal{H}$.

\subsubsection{Development of the game}

Let $U$ be a unit and let $v$ be a vertex. By a slight abuse of notation we will sometimes say "$v$ is added to $U$" to express that $v$ is added to all hyperedges of $U$.

Our goal is to show the following.
\begin{lemm} \label{lemm:NeighbourhoodbMainObser}
There is a hypergraph $\cG$ such that every leaf $u$ of $T_{\cG}$ is the end node of $2 \log d - 6$ bottom units $U_{1}, \ldots, U_{2 \log d - 6}$ such that $l(U_{j}) \leq (1 - c) \log d$ for $j \leq \log d$ and some constant $c > 0$.
\end{lemm}
Before proving Lemma \ref{lemm:NeighbourhoodbMainObser} we show that it implies Lemma \ref{lemm:NeighbourhoodPartbconstructH}. Let $c' = \frac{c}{4}$. 
For each leaf $u$ of $T_{\cG}$ we proceed as follows.
We add two children $v, w$ to $u$ and then for $j = 1 \ldots 2 \log d - 6$ add to $U_{j}$ the node $v$ if $j \leq (1-c')\log d$ and $w$, otherwise. Then the full branch ending at $w$ contains $(1 + c') \log d - 6 \geq (1 + c'') \log d$ bottom units of power 1 for some suitable constant $c'' > 0$. 
Our aim is to apply Observation \ref{obse:NeighbourhoodbMoreGeneralizedsufficientconditiontowin}. (Note that if the full branch ending at $v$ contained the same  amount of bottom units as the full branch ending at $w$ then we would be done.) To this end we will split $v$.
Note that the full branch ending at $v$ has $(1-c') \log d$ units $V_{1}, \ldots, V_{(1-c') \log d}$ of power 1 with $l(V_{j}) = l(U_{j}) + 1 \leq (1 - c) \log d + 1$ for every $j = 1, \ldots, (1-c') \log d$. Since $l(V_{j}) \leq \log d + 1$ we have $|V_{j}| \geq 2$ and therefore every  $V_{j}$ can be partitioned into two units $V'_{j}, V''_{j}$ of power 0 with $|V'_{j}|, |V''_{j}| = \frac{|V_{j}|}{2}$. By applying Corollary \ref{coro:NeighbourhoodbMoreGeneralizedUnitsIncrease} for $i = 2(1-c') \log d$, $r = (1 - c) \log d + 1$ and $s = \log d + 1$ we get that our current hypergraph can be extended at $v$ in such a way that each full branch containing $v$ has $(2 + \frac{c}{2}) \log d$ bottom units.

After repeating this procedure for every leaf $u$ of $T_{\cG}$ we can apply Observation \ref{obse:NeighbourhoodbMoreGeneralizedsufficientconditiontowin}, which completes our proof.

\emph{Proof of Lemma \ref{lemm:NeighbourhoodbMainObser}:} 
For simplicity we assume that $\log d$ is even. We say that a full branch $P$ of a tree $T_{\cal{G}}$ has property $\cal{P}$ if it contains $2 \log d - 6$ bottom units $U_{1}, \ldots, U_{2 \log d - 6}$ such that $l(U_{j}) \leq (1 - c) \log d$ for $j \leq \log d$ and some constant $c > 0$. Our construction of the desired hypergraph $\cG$ will consist of two major steps. The next proposition is about the first step.
\begin{prop} \label{prop:NeighbourhoodbFirstStepToDouble}
 Let $i$ be an integer with $0 \leq i \leq \frac{\log d}{2} - 1$. Let $k_{1} = \log d$, if $i = 0$ and $k_{1} = \log d - i - 2$, otherwise.  
Then there is a hypergraph $\cG$ such that each full branch of $T_{\cG}$ either has property $\cal{P}$ or contains $\log d + i$ bottom units $U_{1}, \ldots, U_{\log d + i}$ with 
 \begin{itemize}
  \item $l(U_{j}) = j$ for $j \leq k_{1}$
  \item $l(U_{k_{1} + 2r - 1}), l(U_{k_{1} + 2r}) = k_{1} + r + 1$  for  $r \geq 1$ 
 \end{itemize}
\end{prop}
\emph{Proof:} We proceed by induction.
By Corollary \ref{coro:NeighbourhoodbAugmentUnits} applied for $i = \log d$ the claim is true for $i = 0$. Suppose that it holds for $i \leq \frac{\log d}{2} - 2$. 
For each leaf $u$ of $T_{\cG}$ we then proceed as follows. If the full branch ending at $u$ has property $\cal{P}$ then we do nothing. Otherwise, induction yields that the full branch ending at $u$ contains $\log d + i$ bottom units $U_{1}, \ldots, U_{\log d + i}$ according to the description in Proposition \ref{prop:NeighbourhoodbFirstStepToDouble}.
We then add two children $v$, $w$ to $u$. For $j = 1 \ldots \log d + i$ we then add to $U_{j}$ the vertex $v$ if $j \leq i + 2$ and $w$, otherwise. 
Note that the full branch ending at $w$ contains $\log d - 2$ bottom units $V_{i + 3}, \ldots, V_{\log d + i}$ of power $1$ with $l(V_{j}) = l(U_{j}) + 1$ for $j = i + 3 \ldots \log d + i$. Since each $V_{j}$ is of length at most $\log d + 1$ it contains at least two hyperedges and can thus be partitioned into two units $V'_{j}, V''_{j}$ of power 0 with $l(V'_{j}), l(V''_{j}) = l(U_{j}) + 1$. Moreover, 
$l(V'_{r}) \leq k_{1} + \lceil \frac{r - k_{1}}{2} \rceil + 1$ (it can be checked that this is true both for $r \geq k_{1}$ and $r \leq k_{1})$. Hence
$l(V'_{i + 2 + \frac{\log d }{2}}) \leq k_{1} + \lceil \frac{i + 2 + \frac{\log d }{2} - k_{1}}{2} \rceil + 1$.
So $l(V'_{i + 2 + \frac{\log d }{2}}) \leq \frac{3}{4} \log d + 1$ and thus the full branch ending at $w$ has property $\cal{P}$.

It remains to consider the full branch $P$ ending at $v$. $P$ contains $i + 2$ units $V_{1}, \ldots, V_{i + 2}$ of power 1, which due to a similar argument as before correspond to $2(i + 2)$ units $V'_{1}, V''_{1}, \ldots, V'_{i + 2}, V''_{i + 2}$ with $l(V'_{j}), l(V''_{j}) = l(U_{j}) + 1 = j  + 1$ (note that $i + 2 \leq k_{1}$) for $j = 1 \ldots i + 2$. By applying Corollary \ref{coro:NeighbourhoodbMoreGeneralizedUnitsIncrease} for $r = i + 3$ and  $s = \log d$ we get that our current hypergraph can be extended at $v$ in such a way that each full branch containing $v$ has the $\log d + i + 1$ required bottom hyperedges (considering the induction hypothesis for $i + 1$). 
After repeating this procedure for every leaf $u$ of $T_{\cal{G}}$ the resulting hypergraph fulfills our hypothesis for $i + 1$. $\Box$  

The following corollary specifies the result of our first step.
\begin{coro} \label{coro:NeighbourhoodbFirstDoubleStepResult}
 Let $k_{1} = \frac{\log d}{2} - 1$. Then there is a hypergraph $\cal{G}$ such that each full branch of $T_{\cG}$ either has property $\cal{P}$ or contains $\frac{3}{2} \log d - 1$ units $U_{1}, \ldots, U_{\frac{3}{2} \log d - 1}$ such that  
\begin{itemize}
  \item $l(U_{j}) = j$ for $j \leq k_{1}$
  \item $l(U_{k_{1} + 2r - 1}), l(U_{k_{1} + 2r}) = k_{1} + r + 1$ for $r \geq 1$ 
 \end{itemize}
 \end{coro}
The next proposition deals with the second major step of our construction.
 \begin{prop} \label{prop:NeighborhoodDegreeRegularSecondStep}
  Let $i$ be an integer with $\frac{\log d}{2} - 1 \leq i \leq \log d - 6$ and let 
  \newline
  $k_{1} = \frac{\log d}{2} - 1$, if $i = \frac{\log d}{2} - 1$ and $k_{1} = \log d - i - 4$, otherwise. Then there is a $k_{2} \geq 2$ such that there is a hypergraph $\cG$ where each full branch of $T_{\cal{G}}$ either has property $\cal{P}$ or contains $\log d + i$ units $U_{1}, \ldots, U_{\log d + i}$ with
  \begin{itemize}
  \item $l(U_{j}) \leq j$ for $j \leq k_{1}$
  \item $l(U_{k_{1} + 2r - 1}), l(U_{k_{1} + 2r}) \leq k_{1} + r + 1$ for $1 \leq r \leq k_{2}$ 
  \item $l(U_{k_{1} + 2k_{2} + 2m - 1}), l(U_{k_{1} + 2k_{2} + 2m}) \leq k_{1} + k_{2} + m + 2$ for $m \geq 1$
 \end{itemize}
\end{prop}
Note that Proposition \ref{prop:NeighborhoodDegreeRegularSecondStep} applied for $i = \log d - 6$ directly implies Lemma \ref{lemm:NeighbourhoodbMainObser}. $\Box$

So it remains to show Proposition \ref{prop:NeighborhoodDegreeRegularSecondStep}.
\newline
\emph{Proof of Proposition \ref{prop:NeighborhoodDegreeRegularSecondStep}:} Corollary \ref{coro:NeighbourhoodbFirstDoubleStepResult} yields that our claim is true for $i = \frac{\log d}{2} - 1$ (with $k_{2} = \infty$). Suppose that the claim holds for $i$. 
For each leaf $u$ of $T_{\cG}$ we proceed as follows. If the full branch ending at $u$ has property $\cal{P}$ we do nothing. Otherwise induction yields that the full branch ending at $u$ contains $\log d + i$ bottom units $U_{1}, \ldots, U_{\log d + i}$ according to the description in Proposition \ref{prop:NeighborhoodDegreeRegularSecondStep}.
In this case we add two children $v$, $w$ to $u$ and for $j = 1 \ldots \log d + i$ add to $U_{j}$ the node $v$, if $j \leq i  + 3$ and $w$, otherwise. 
The full branch $P$ ending at $w$ contains $\log d - 3$ units $U'_{i + 4}, \ldots U'_{\log d + i}$ of power 1 with $l(U'_{j}) = l(U_{j}) + 1$. The induction hypothesis yields that for each $U'_{j}$ we have $l(U'_{j}) \leq \log d$, implying that $|U'_{j}| \geq 2$. So $U'_{j}$ can be partitioned into two units $V'_{j}, V''_{j}$ of power 0 with $l(V'_{j}), l(V''_{j}) = l(U'_{j})$.
Due to our hypothesis $l(V'_{j})$ (and $l(V''_{j})$, respectively) is at most $k_{1} + 2 + \lceil \frac{j - k_{1}}{2} \rceil$ (note that this also holds for $j \leq k_{1}$) and so for $j$ with $i + 4 \leq j \leq i + 3 + \frac{\log d}{2}$ we have $l(V'_{j}) \leq \frac{k_{1}}{2} + 3 + \frac{i + 3}{2} + \frac{\log d}{4} \leq \frac{3}{4} \log d + 3$. 
Since $P$ contains $V'_{i + 4}, V''_{i + 4}, \ldots, V'_{\log d + i}, V''_{\log d + i}$ it has property $\cal{P}$.

It remains to consider the full branch $P$ ending at $v$. $P$ contains $i + 3$ units $U'_{1} \ldots U'_{i + 3}$ of power 1. For a similar reason as above 
they can be partitioned into $2(i + 3)$ units $V^{(1)}_{1}, V^{(2)}_{1}, \ldots V^{(1)}_{i + 3}, V^{(2)}_{i + 3}$ with $l(V^{(s)}_{j}) = l(U_{j}) + 1$ for $s \in \{1,2\}$. According to our assumption we have  for $s \in \{1,2\}$
\begin{itemize}
 \item $l(V^{(s)}_{j}) \leq j + 1$ for $j \leq k_{1}$
 \item $l(V^{(s)}_{k_{1} + 2r - 1}), l(V^{(s)}_{k_{1} + 2r}) \leq k_{1} + r + 2$ for $1 \leq r \leq k_{2}$
 \item $l(V^{(s)}_{k_{1} + 2k_{2} + 2m - 1}), l(V^{(s)}_{k_{1} + 2k_{2} + 2m}) \leq k_{1} + k_{2} + m + 3$ for $m \geq 1$
\end{itemize}
Note that for each $V^{(s)}_{j}$ 
we have $l(V^{(s)}_{j}) \leq j + 2 \leq i + 5$ (this can be seen by considering each of the three possible intervals for $j$ separately and using that $k_{2} \geq 1$).
Let $k'_{1} = \log d - i - 5$. By applying Corollary \ref{coro:NeighbourhoodbMoreGeneralizedUnitsIncrease} for $r = i + 5$ and $s = \log d$ we obtain that our current graph can be extended at $v$ in such a way that each full branch of the tree $T_{\cal{G'}}$ of the resulting graph $\cal{G'}$ contains
$\log d + i + 1$ units $X_{1}, \ldots, X_{k'_{1}}$, $W^{(1)}_{1}, W^{(2)}_{1}, \ldots, W^{(1)}_{i + 3}, W^{(2)}_{i + 3}$ with
\begin{itemize}
 \item $l(X_{j}) \leq j$ for $j \leq k'_{1}$
 \item $l(W^{(s)}_{j}) \leq j + k'_{1} + 1$ for $s \in \{1,2\}$ and $j \leq k_{1} $
 \item $l(W^{(s)}_{k_{1} + 2r - 1}), l(W^{(s)}_{k_{1} + 2r}) \leq k_{1} + k'_{1} + r + 2$ for  $s \in \{1,2\}$ and  $r \leq k_{2}$
 \item $l(W^{(s)}_{k_{1} + 2k_{2} + 2m - 1}), l(W^{(s)}_{k_{1} + 2k_{2} + 2m}) \leq k_{1} + k_{2} + k'_{1} + m + 3$ for  $s \in \{1,2\}$ and $m \geq 1$
\end{itemize}
Let $i' = i + 1$ and $k'_{2} = k_{1}$. Note that $k'_{1} = \log d - i' - 4$ and that $k'_{2} \geq 2$ (due to the fact that by definition $k_{1} \geq 2$). The fact that $k_{2} \geq 2$ guarantees that after a suitable renaming the units $X_{1}, \ldots, X_{k'_{1}}$, $W^{(1)}_{1}, W^{(2)}_{1}, \ldots, W^{(1)}_{i + 3}, W^{(2)}_{i + 3}$ fulfill our hypothesis for $i', k'_{1}$ and $k'_{2}$. $\Box$

\section{Establishing a Proper Halving 2-Coloring}

\emph{Proof of Theorem \ref{theo:NeighbourhoodPartProperHalvingIsOk}:} For simplicity we only consider hypergraphs with an even number of vertices. We will show the following stronger claim. 
\begin{prop} \label{prop:NeighbourhoodPartasworksforhalvingstronger}
Let $\cal{F}$ be a $n$-uniform hypergraph with maximum degree at most $\frac{2^{n}}{4en}$. Then for each pairing $(v_{i_{1}}, w_{i_{1}}), (v_{i_{2}}, w_{i_{2}}), (v_{i_{3}}, w_{i_{3}}), \ldots $ of the vertices of $\cal{F}$ there is a proper 2-coloring such that $v_{i_{k}}$ and $w_{i_{k}}$ have different colors for each $k$.
\end{prop}
To prove Theorem \ref{theo:NeighbourhoodPartProperHalvingIsOk} it suffices to prove Proposition \ref{prop:NeighbourhoodPartasworksforhalvingstronger}. We adapt a proof by Kratochv\'{i}l, Savick\'y and Tuza \cite{KST} .
\newline
\emph{Proof of Proposition \ref{prop:NeighbourhoodPartasworksforhalvingstronger}:} 
Our claim is a consequence of Lov\'{a}sz Local Lemma.
\begin{lemm} {\bf{(Lov\'{a}sz Local Lemma.)}}
Let $A_{1}, \ldots, A_{m}$ be events in some probability space, and let $G$ be a graph with vertices $A_{1}, \ldots, A_{m}$ and edges $E$ such that each $A_{i}$ is mutually independent of all the events $\{A_{j} \enspace | \enspace \{A_{i}, A_{j}\} \notin E, i \neq j\}$. If there exist real numbers $0 < \gamma_{i} < 1$ for $i = 1, \ldots, m$ satisfying
\begin{displaymath}
\text{Pr}(A_{i}) \leq \gamma_{i}  \prod_{j: (A_{i}, A_{j}) \in E} (1 - \gamma_{i})
\end{displaymath}
for all $i = 1, \ldots, m$ then
\begin{displaymath}
\text{Pr}(\neg A_{1} \wedge \neg A_{2} \wedge \cdots \wedge \neg A_{m}) > 0
\end{displaymath}
\end{lemm}
For a proof of the Lov\'{a}sz Local Lemma and different versions, see e.g. \cite{AS}.
Let $d = \frac{2^{n}}{4en}$. Note that each proper coloring of $\cal{F}$ fulfilling the condition that  $v_{i_{k}}$ and $w_{i_{k}}$ have different colors for each $k$ is a proper-2-coloring.
In each edge of $\cal{F}$ we then replace $w_{i_{k}}$ with $\bar{v_{i_{k}}}$, expressing that  $w_{i_{k}}$ gets the "inverse" color of $v_{i_{k}}$. Let $\cal{F'}$ denote the resulting hypergraph. Note that the maximum degree of $\cal{F'}$ is at most $2d = \frac{2^{n}}{2en}$. Indeed, the degree of $v_{i_{k}}$ is bounded by the number of  edges possessing $v_{i_{k}}$ plus the number of edges possessing $\bar{v_{i_{k}}}$. Since edges containing both $v_{i_{k}}, \bar{v_{i_{k}}}$ get two colors in every coloring we can ignore those edges and assume that no edge of $\cal{F'}$ contains both $v_{i_{k}}, \bar{v_{i_{k}}}$ for some $k$.
Since every proper 2-coloring of $\cal{F'}$ directly provides the desired proper halving 2-coloring. it suffices to show that $\cal{F'}$ has a proper 2-coloring. To this end we apply the Lov\'{a}sz Local Lemma. Let the probability space be the set of all color assignments to the vertices of $\cal{F}$ with the uniform distribution. Let $E({\cal{F'}}) = \{ E_{1}, \ldots, E_{m} \}$ and let $A_{i}$ be the event that $E_{i}$ is monochromatic in a random 2-coloring. Let $G$ be the graph where $A_{i}$ and $A_{j}$ are connected if they have a vertex in common. Since every vertex has degree at most $2d$ every $A_{i}$ has degree at most $n \cdot (2d - 1)$. 
Note that $\text{Pr}(A_{i} = 1) = 2 \cdot 2^{-n}$. We let $\gamma_{i} = e \cdot \text{Pr}(A_{i} = 1) = 2e \cdot 2^{-n}$ for each $i$. Hence
\begin{displaymath}
\frac{\gamma_{i}}{\text{Pr}(A_{i} = 1)} \prod_{A_{i}, A_{j} \in E(G)} (1 - \gamma_{j}) \geq e \left(1 - \frac{2e}{2^{n}} \right)^{n \left(\frac{2^{n}}{2en} - 1 \right)} > e \left(1 - \frac{2e}{2^{n}}\right)^{\frac{2^{n}}{2e} - 1} > ee^{-1} = 1
\end{displaymath}
Hence $Pr(\neg A_{1} \wedge \neg A_{2} \wedge \cdots \wedge \neg A_{m}) > 0$ and therefore there is a proper 2-coloring on $\cal{F'}$. $\Box$


\end{document}